\documentclass[%
 reprint,
 amsmath,amssymb,
 aps,
prb,
]{revtex4-1}

\usepackage{graphicx}
\usepackage{dcolumn}
\usepackage{bm}
\usepackage[colorlinks=true, citecolor=red, linkcolor=blue, urlcolor=black]{hyperref}

\def\d{{\textit{d}}}

\newcommand{\bra}[1]{\langle#1\rvert}
\newcommand{\ket}[1]{\lvert#1\rangle}
\newcommand{\mean}[1]{\langle#1\rangle}

\newcommand{\Bra}[1]{\left\langle#1\left\rvert}
\newcommand{\Ket}[1]{\right\lvert#1\right\rangle}

\def\tjz{{$t$--$J^z$}}

\begin{document}


\title{Hole in the 2D Ising Antiferromagnet: Origin of the Incoherent Spectrum}

\author{Piotr Wrzosek$^1$}
\email{Piotr.Wrzosek@fuw.edu.pl}
\author{Krzysztof Wohlfeld$^1$}
 
\affiliation{%
$^1$Institute of Theoretical Physics, Faculty of Physics, University of Warsaw, Pasteura 5, PL-02093 Warsaw, Poland
}%

\date{\today}

\begin{abstract}
We develop a `self-avoiding walks' approximation and use it to calculate the spectral function of a single hole introduced into the 2D square lattice Ising antiferromagnet. The obtained local spectral function qualitatively agrees with the exact diagonalisation result and is largely incoherent. Such a result stays in contrast with the spectrum obtained on a Bethe lattice, which consists of the well-separated quasiparticle-like peaks and stems from the motion of a hole in an effective linear potential.  We determine that this onset of the incoherent spectrum on a square lattice (i) is not triggered by the so-called Trugman loops but (ii) originates in the warping of the linear potential by the interactions between magnons created along the tangential paths of the moving hole.
\end{abstract}

\pacs{Valid PACS appear here}
\maketitle


\section{\label{sec:intro}Introduction}

\begin{figure}[b!]
	\includegraphics[width=0.48\textwidth]{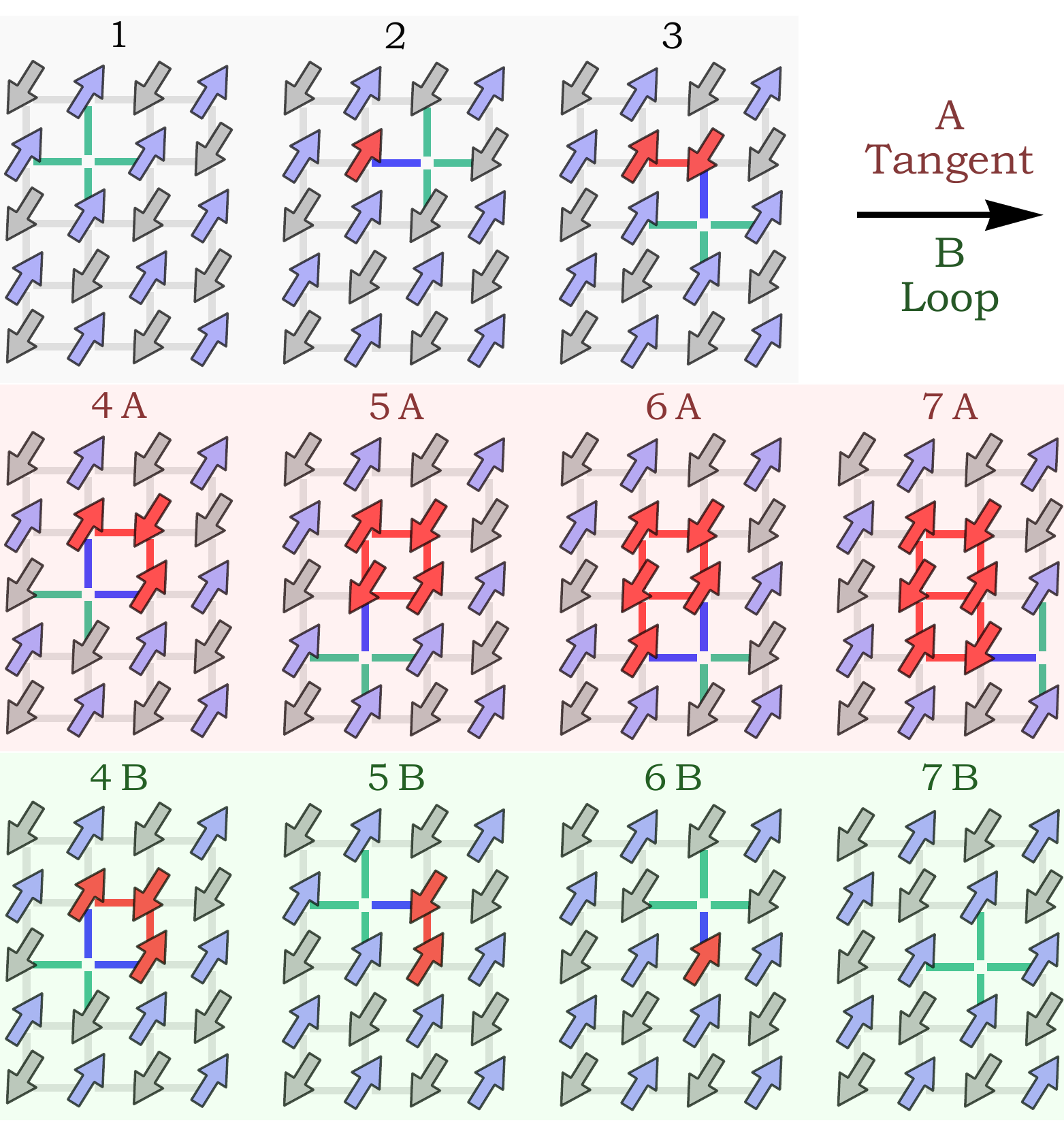}
	\caption{
Cartoon picture of the motion of a hole (due to tunneling of electrons) in the ground state of the Ising antiferromagnet on a square lattice. Two antiferromagnetic sublattices are colored in light blue and gray; electrons may jump over blue and green bonds; red arrows depict spins with unsatisfied bonds and misaligned w.r.t the ground state as a result of the hole motion. Top panels: creation of one misaligned spin at each hole hop. Middle and bottom panels: a partial or full reconstruction of the antiferromagnetic order {\it either} by moving along a path tangential to itself (subfigures $4A\rightarrow 5A \rightarrow 6A \rightarrow 7A$; as discussed in this paper) {\it or} by going around a loop (panels $4B\rightarrow 5B \rightarrow 6B \rightarrow 7B$; as discussed by Trugman in~\onlinecite{Tru88}). 
	}
	\label{fig:cartoon}
\end{figure}

The problem of a single hole doped into the Ising antiferromagnet, with its dynamics governed by the $t$--$J^z$ model~\cite{Khomskii2010}, is one of the oldest problems of correlated electron systems. The basic physics of this problem was already understood about 50 years ago~\cite{Bul68}: It is based on the idea that the hole is subject to an effective potential originating from the energy cost associated with the antiferromagnetic bonds being gradually destroyed by the mobile hole. As the number of the broken bonds, originating in the `misalignment' of spins, cf. top panels of ~Fig.~\ref{fig:cartoon}, is assumed to be the same at each time the hole hops between the nearest neighbour sites the effective potential grows linearly with the distance covered by the mobile hole. Consequently, the corresponding spectral function of this problem is `ladder-like’, i.e. it consists of the well-separated quasiparticle-like peaks with the low-lying ones split by a gap $\propto (J^z/t)^{2/3}$.~\cite{Bul68, Kan89, Mar91}
The above picture was qualitatively confirmed by a number of works, for example by applying the retraceable path approximation~\cite{Bri70} to this problem~\cite{Bul68, Shr88}, using the self-consistent Born approximation~\cite{Sch88, Kan89, Mar91} whose equations can be analytically written using a closed form~\cite{Sta96}, extending the latter one to include the magnon-magnon interactions~\cite{Che99, Bie19}, or designing the so-called magnon expansion method~\cite{Bie19}. 

To the best of our knowledge, the only qualitatively different scenario to the above picture came from the paper by Trugman~\cite{Tru88}. There it was suggested that the hole may go along a loop (hereafter called Trugman), reverse the misaligned spins and thus `liberate itself' from the linear potential, cf.~bottom panels of ~Fig.~\ref{fig:cartoon}. It turned out that such a process lead to a momentum-dependence of the spectral function, though for instance its contribution to the ground state energy is estimated to be relatively small~\cite{Bar89, Che99, Bie19}. Indeed the magnon expansion results~\cite{Bie19}, which include the Trugman loops, advocate that the ladder-like spectrum should not be substantially affected by the closed loop processes. Interestingly, Macready and Jacobs~\cite{Mac91} consider the \tjz~model in the hopping basis and obtain a result that could suggest a breakdown of the ladder spectrum in the \tjz~model on a square lattice. However, they do not comment on this point.

Thus, one could think of this problem as solved and completely understood. However, a closer look at the spectral functions of the half-filled $t$--$J^z$ model calculated using exact diagonalisation (ED), cf. Fig.~4 of Ref.~\onlinecite{Mar91} (or cf. Fig.~\ref{fig:1} below), suggests that the spectrum does not at all seem to be ladder-like. Note that such an incoherent spectrum is rather not an artefact of the ED method. In fact, the spectrum of a finite system considered by ED would look {\it more} coherent than the one of the infinite system, for the former naturally forms a discrete spectrum.

In order to resolve the above question we concentrate our attention at a particular physical process caused by the motion of the hole along the so-called `tangential paths', as pictorially shown in the middle panels of ~Fig.~\ref{fig:cartoon}. Intuitively, a tangential path is a path along which the hole has moved such that the path touches `itself', cf. middle panels of Fig~\ref{fig:cartoon}. Formally, the tangential path is a path which includes at least one {\it pair} of distinct lattice sites. These lattice sites fulfill the following conditions: (i) They are nearest neighbors; (ii) They belong to the path along which the hole has moved (i.e. they have been visited by the hole); (iii) The hole has not moved along the bond connecting the two sites forming the pair. As a side note, we stress that a set of paths containing loops and the set of paths containing tangents are {\it not} disjoint, i.e. there can be a path which contains loops and has also tangential segments. 

In this paper we show that including the tangential paths leads to the onset of a largely incoherent spectral function for higher energies and explains the difference between the ED spectra and all the other approaches. This is due to the warping of the linear potential caused by the distinct energy costs associated with the misaligned spins along the tangential paths. To this end we solve the model using a self-avoiding walks approximation, a semi-analytic approach which neglects all walks with loops but otherwise is exact, i.e. in particular it properly includes all tangential paths without loop segments. Crucially, despite overlooking all loop paths, we show that such an approximation reproduces surprisingly well the ED spectrum.

As explained in detail in the paper such a mechanism works only when the proper geometry of the lattice is taken into account and requires solving the `full' $t$--$J^z$ model. In other words, approximating the square lattice by a Bethe lattice with an appropriate coordinate number {\it or} performing the linear spin wave approach (i.e. neglecting the magnon-magnon interaction) would not lead to the collapse of the ladder spectrum. Thus, the presented here mechanism was not discussed in several works mentioned above, for they either were based on the Bethe lattice geometry or relied on the linear spin wave theory approximation~\footnote{Or, in case of~\cite{Bie19}, the number of magnons was not enough to properly calculate the energies of the high energy states}.

The paper is organised as follows. In Sec.~\ref{sec:model} we write down the \tjz~Hamiltonian and map it onto the polaronic Hamiltonian using the slave-fermion transformation. In Sec.~\ref{sec:methods} we describe the self-avoiding walks approximation method. In Sec.~\ref{sec:results} the spectra obtained using this method are successfully benchmarked against the ED results on a finite cluster. The origin of such a good agreement is attributed to the low significance of the (Trugman) loops in the considered here regime of $J/t \in [0.4, 2]$, as discussed in detail in Sec.~\ref{sec:agr}. Finally, in Sec.~\ref{sec:discussion} we explain the origin of the largely incoherent spectra obtained in Sec.~\ref{sec:results} as stemming from the warping of the linear potential due to the interactions between magnons along the tangential paths taken by the moving hole (we also show in Appendix~\hyperref[appendix]{A} that such an incoherent spectrum is not triggered by an apparent superposition of the coherent momentum-dependent spectral functions). We conclude the paper in Sec.~\ref{sec:conclusions}.

\section{\label{sec:model}Model:\\* from $t$--$J^z$ to polaronic Hamiltonian}

\subsection{$t$--$J^z$ model}
The Hamiltonian of the \tjz~model~\cite{Khomskii2010},
\begin{equation}\label{eq:H}
\mathcal{H} = -t \sum_{\mean{i,j},\sigma} \left( \tilde{c}_{i\sigma}^\dag \tilde{c}_{j\sigma} + \text{H.c.} \right) + J \sum_{\mean{i,j}} \left( S_i^z S_j^z - \frac{1}{4} \tilde{n}_i \tilde{n}_j \right),
\end{equation}
describes energy of the system of the `constrained electrons' $\tilde{c}_{i\sigma}^\dag = c_{i\sigma}^\dag(1 - n_{i\bar{\sigma}})$ tunnelling with amplitude $t$ to the nearest unoccupied lattice sites. Note that if two nearest neighbour sites are occupied by electrons, then tunnelling is impossible (hence the term `constrained electrons') but instead the $z$ components of the spins $S_i^z$ carried by the electrons interact with an exchange constant $J$.

In the half-filled limit each site is occupied by exactly one electron and for $J > 0$ the ground state of the system is an Ising antiferromagnet. Here we consider a single hole injected into the ground state of the half-filled \tjz~model. The quantity of interest is the {\it local} Green's function,
\begin{equation}\label{eq:Gsigma}
G_\sigma(\omega) = \Bra{\textsc{gs}} \tilde{c}_{i\sigma}^\dag \frac{1}{\omega - \mathcal{H} + E_\textsc{gs}} \tilde{c}_{i\sigma} \Ket{\textsc{gs}},
\end{equation}
where $E_\textsc{gs}$ is the energy of the ground state (GS). From the above Green's function we calculate the local spectral function defined as,
\begin{equation}
A_\sigma(\omega) = -\frac{1}{\pi}\lim_{\delta\to0^+} \text{Im}\{G_\sigma(\omega + i\delta)\},
\end{equation}
which is the central object in this paper. In what follows we are interested in the local spectral function $A(\omega)$ calculated for two different lattice geometries: the Bethe lattice with the coordinate number $z$ and the square lattice.

\subsection{Slave-fermion transformation}
Let us reformulate the stated problem in the form of a polaronic Hamiltonian, i.e. we express the (constrained) electron and spin operators in terms of the (fermionic) hole $h_i^\dag$ and (bosonic) magnon $a_i^\dag$ operators, cf.~\cite{Kan89, Mar91, Sta96, Che99, Bie19}. To this end, we first split the lattice into two sublattices $A$ and $B$, each consisting respectively of spins up and down in the ground state. Next, without loss of generality we rotate all spins on sublattice B,
\begin{equation}\label{eq:sf1}
\forall_{j \in {\rm B}}\quad  S^z_j \rightarrow -S^z_j.
\end{equation}
Finally, we introduce the hole and magnon operators in terms of the following slave-fermion transformations,
\begin{equation}\label{eq:sf2}
\begin{aligned}
\tilde{c}_{i\uparrow}^\dag &= h_i, &\quad \tilde{c}_{i\uparrow} &= h_i^\dag (1 - a_i^\dag a_i), \\
\tilde{c}_{i\downarrow}^\dag &= h_i a_i^\dag, &\quad \tilde{c}_{i\downarrow} &= h_i^\dag a_i,
\end{aligned}
\end{equation}
\begin{equation}\label{eq:sf3}
\begin{aligned}
S_i^z &= \frac{1}{2} - a_i^\dag a_i - \frac{1}{2}h_i^\dag h_i, \\
\tilde{n}_i &= 1 - h_i^\dag h_i.
\end{aligned}
\end{equation}

Let us stress that the above slave-fermion transformation is very common to the `single hole in the antiferromagnet' problems~\cite{Kan89, Mar91, Sta96, Che99, Bie19}, although note that, unlike e.g. in~\cite{Kan89, Mar91}, below magnons are not subject to the linear spin wave approximation. Thus, the eigenstates of the resulting polaronic Hamiltonian (see below) are exactly the same as of the ones of the original \tjz~Hamiltonian. 

\subsection{Polaronic model}
Applying the slave-fermion transformation (\ref{eq:sf1}-\ref{eq:sf3}) to Hamiltonian \eqref{eq:H} leads to
\begin{align}
\mathcal{H} &= \mathcal{H}_t + \mathcal{H}_J,
\end{align}
where the kinetic energy in terms of hole and magnon operators reads,
\begin{equation}
\begin{split}
\mathcal{H}_t = &-t \sum_{\mean{i,j}} h_i^\dag h_j \left( a_i + a_j^\dag (1 -  a_i^\dag a_i) \right) \\
&-t \sum_{\mean{i,j}} h_j^\dag h_i \left( a_j + a_i^\dag (1 -  a_j^\dag a_j) \right).
\end{split}
\label{eq:ht}
\end{equation}
and the potential energy of the system reads,
\begin{equation}
\begin{aligned}
\mathcal{H}_J = E_\textsc{gs} &+ \frac{J}{2} \sum_{\mean{i,j}} \left( h_i^\dag h_i + h_j^\dag h_j + a_i^\dag a_i + a_j^\dag a_j \right) \\
&- \frac{J}{2} \sum_{\mean{i,j}} \left( h_i^\dag h_i h_j^\dag h_j + 2 a_i^\dag a_i a_j^\dag a_j \right) \\
&- \frac{J}{2} \sum_{\mean{i,j}} \left( h_i^\dag h_i a_j^\dag a_j + h_j^\dag h_j a_i^\dag a_i \right).
\end{aligned}
\label{eq:hj}
\end{equation}
The rotation on the ground state leads to a state that has all the spins pointing up, i.e. $\ket{\varnothing} = \prod_i \tilde{c}_{i\uparrow}^\dag \ket{\varnothing_\text{e}}$, where $\ket{\varnothing_\text{e}}$ is a vacuum state for electrons. Note that then $\ket{\varnothing}$ is a vacuum state for both holes and magnons. This comes from the transformation we use. There are no magnons in $\ket{\varnothing}$ since we associate magnons only with spins pointing down $(\tilde{c}_{i\downarrow}^\dag = h_i a_i^\dag)$ and there are no holes in $\ket{\varnothing}$ since we annihilate all of them $(\tilde{c}_{i\uparrow}^\dag = h_i)$ starting from $\ket{\varnothing_\text{e}}$ which is fully occupied by holes. Then we can define a state with a single hole at site $i$ as,
\begin{equation}
\ket{\psi_0} \equiv h_i^\dag \ket{\varnothing} = h_i^\dag (1 - a_i^\dag a_i) \ket{\varnothing} = \tilde{c}_{i\uparrow} \ket{\varnothing}.
\end{equation}
The local Green's function of a single hole may be therefore written in the following way,
\begin{equation}
G(\omega) = \bra{\psi_0} \hat{G} \ket{\psi_0},
\end{equation} 
where 
\begin{align}\label{eq:G}
\hat{G} = (\omega - \mathcal{H} + E_\textsc{gs})^{-1}. 
\end{align}
Crucially, by comparing Eq.~\eqref{eq:G} with Eq.~\eqref{eq:Gsigma} we observe that either $G_\sigma(\omega) = 0$ or $G_\sigma(\omega) = G(\omega)$. The same applies to the local spectral function,
\begin{equation}
A(\omega) = -\frac{1}{\pi}\lim_{\delta\to0^+} \text{Im}\{G(\omega + i\delta)\},
\end{equation}
i.e. either $A_\sigma(\omega) = 0$ or $A_\sigma(\omega) = A(\omega)$.

\begin{figure*}[t]
  \includegraphics[width=0.98\textwidth]{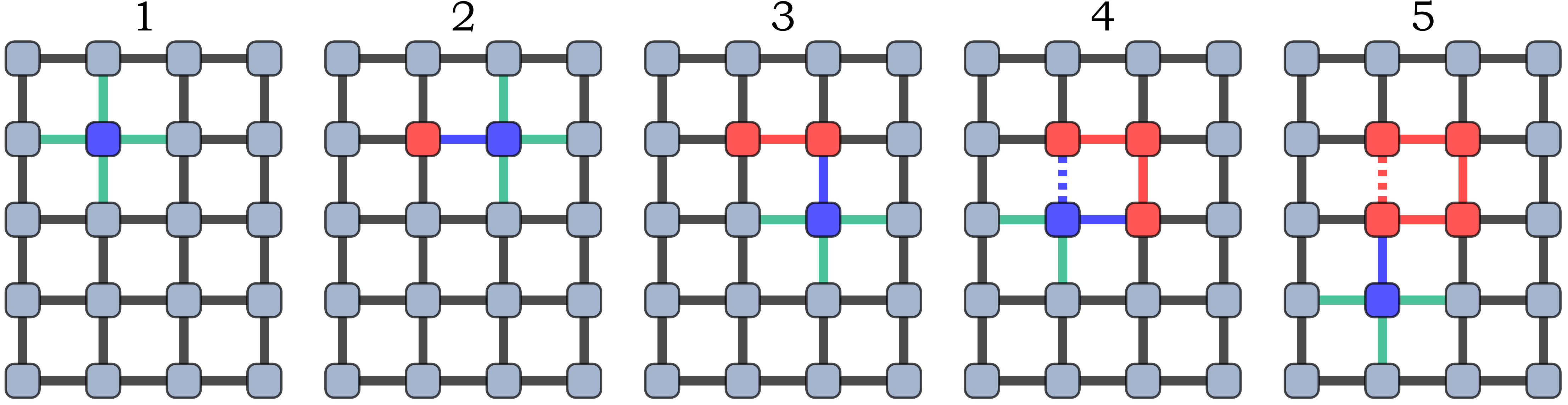}
  \caption{
Cartoon picture of the hole motion on a square lattice in terms of holes $(h_i^\dag h_i)$ and magnons $(a_i^\dag a_i)$. Blue square represents the hole, red squares represent magnons created by the hole, gray squares represent empty sites. Within the self-avoiding walks approximation propagation via the dashed bond is forbidden. In addition, subfigures 4 and 5 show `satellite' hole-magnon proximity interaction (blue dashed bond) and `satellite' magnon-magnon interaction (red dashed bond), both are not possible on the Bethe lattice (cf. FIG. \ref{fig:bethe}). See main text for further details.
}
\label{fig:square}
  \includegraphics[width=0.98\textwidth]{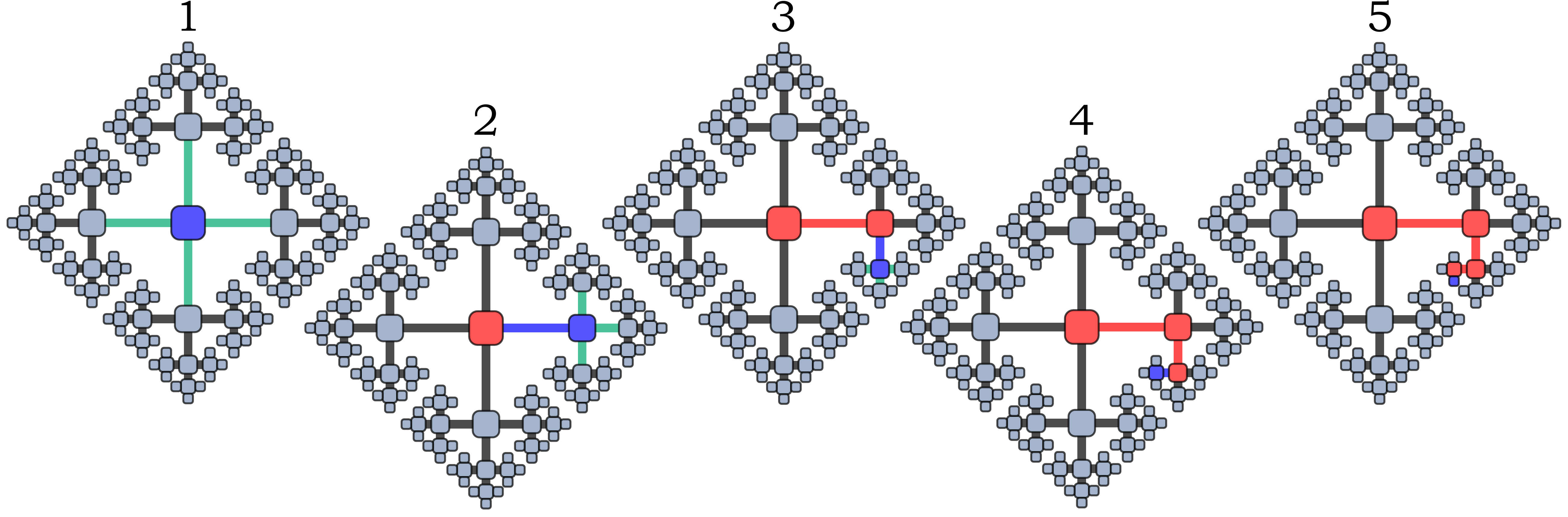}
  \caption{
Cartoon picture of the hole motion on a Bethe lattice with the coordinate number $z = 4$ in terms of holes $(h_i^\dag h_i)$ and magnons $(a_i^\dag a_i)$. Symbols are the same as in the FIG. \ref{fig:square}. Note the lack of the `satellite' bonds (no dashed bonds) in contrast to FIG. \ref{fig:square}. See main text for further details.
}
\label{fig:bethe}
\end{figure*}

It is instructive to compare the propagation of a single hole in the language of the original \tjz~and the obtained polaronic Hamiltonian. To this end, we plot a cartoon picture of the hole propagation on a square lattice in the polaronic language, cf. Fig~\ref{fig:square}, which exactly `mimics' the hole propagation shown in the \tjz~model language on the five subfigures of Fig.~\ref{fig:cartoon}.  (Note that Fig.~\ref{fig:bethe} shows the same propagation as Fig.~\ref{fig:square} but on a Bethe lattice---the differences between the two are discussed in the next section.) One can see that the propagating hole either creates a magnon or annihilates one, depending on whether the site to which the hole propagates contains a magnon or not, as described by~\eqref{eq:ht}. Such a hole motion leads to the changes in the potential energy associated with the cost of having a magnon in the system ($a_i^\dag a_i$), which may be further affected by the magnon-magnon interaction terms ($a_i^\dag a_i a_j^\dag a_j$) and the hole-magnon proximity interaction terms ($h_i^\dag h_i a_j^\dag a_j$), cf.~\eqref{eq:hj}. Note that whereas adding a single hole always costs energy $\propto J$ due to the $h_i^\dag h_i$ terms in~\eqref{eq:hj}, the hole-hole interaction $(h_i^\dag h_i h_j^\dag h_j)$, also present in \eqref{eq:hj}, does not play any role in the case of a single hole.

\section{\label{sec:methods}Methods:\\* Self-avoiding walks approximation}
The central goal of this paper is to calculate the above-defined local spectral function $A(\omega)$ on the square as well as on the Bethe lattice. The subspace $\mathcal{S}$ of all states reachable from the initial state $\ket{\psi_0}$ through operator $\hat{G}$ can be obtained in the same way for both lattices of interest provided that in case of square lattice we restrict the Hilbert space to the states reachable by the so-called self-avoiding walks~\cite{saw_lectures} (hence the name: self-avoiding walks approximation). For the square lattice this means that propagation via the dashed bonds presented in Fig.~\ref{fig:square} is forbidden, for the hole is not allowed to cross its own path (i.e. the walk is self-avoiding). Thus, all hops along the closed (Trugman) loops are naturally excluded in this approximation. We note that, `physically', i.e. from the point of Hamiltonian \eqref{eq:H} defined on a square lattice, there is no difference between the `regular' vs. the `satellite' bonds (i.e. the red / blue solid vs. dashed bonds of Fig.~\ref{fig:2}). The distinction between these two types of bonds is due to  the self-avoiding walks approximation and thus it is this approximation which defines the `satellite' bonds.

On the other hand, the Bethe lattice is a tree (i.e.~a~connected acyclic graph). Thus, all walks form the one-dimensional-like chains and by definition the walk on the Bethe lattice is self-avoiding, cf. the lack of the dashed bonds in Fig~\ref{fig:bethe}. 

We start by introducing the spanning operator $\mathcal{A}^\dag$,
\begin{align}
\mathcal{A}^\dag \ket{\psi} = \bigcup_{\mean{i,j}}&~\left\{ h_i^\dag h_j a_j^\dag (1 -  a_i^\dag a_i) \ket{\psi} \right. \nonumber  \\ &+ \left. h_j^\dag h_i a_i^\dag (1 -  a_j^\dag a_j) \ket{\psi} \right\} \setminus \{0\},
\end{align}
and we divide subspace $\mathcal{S}$ into the subsets consisting of states with given number of magnons $n$,
\begin{equation} \label{eq:S_def}
\mathcal{S} = \bigcup_{n} \mathcal{S}_n,
\end{equation}
where $\mathcal{S}_0 = \{\ket{\psi_0}\}$ and $\mathcal{S}_n = \bigcup_{\ket{\psi} \in \mathcal{S}_{n-1}} \mathcal{A}^\dag \ket{\psi}$. We also denote the index set of $S_n$ as $I_n$. As already mentioned, the above-defined set $\mathcal{S}$ of reachable states includes all states reachable on the Bethe lattice which allows us to calculate $G(\omega)$ exactly in this case. In the square lattice case it includes all possible states that can be obtained provided that the hole does not cross its own path.

\subsection{Bethe lattice}
\begin{figure}[t!]
  \includegraphics[width=0.32\textwidth]{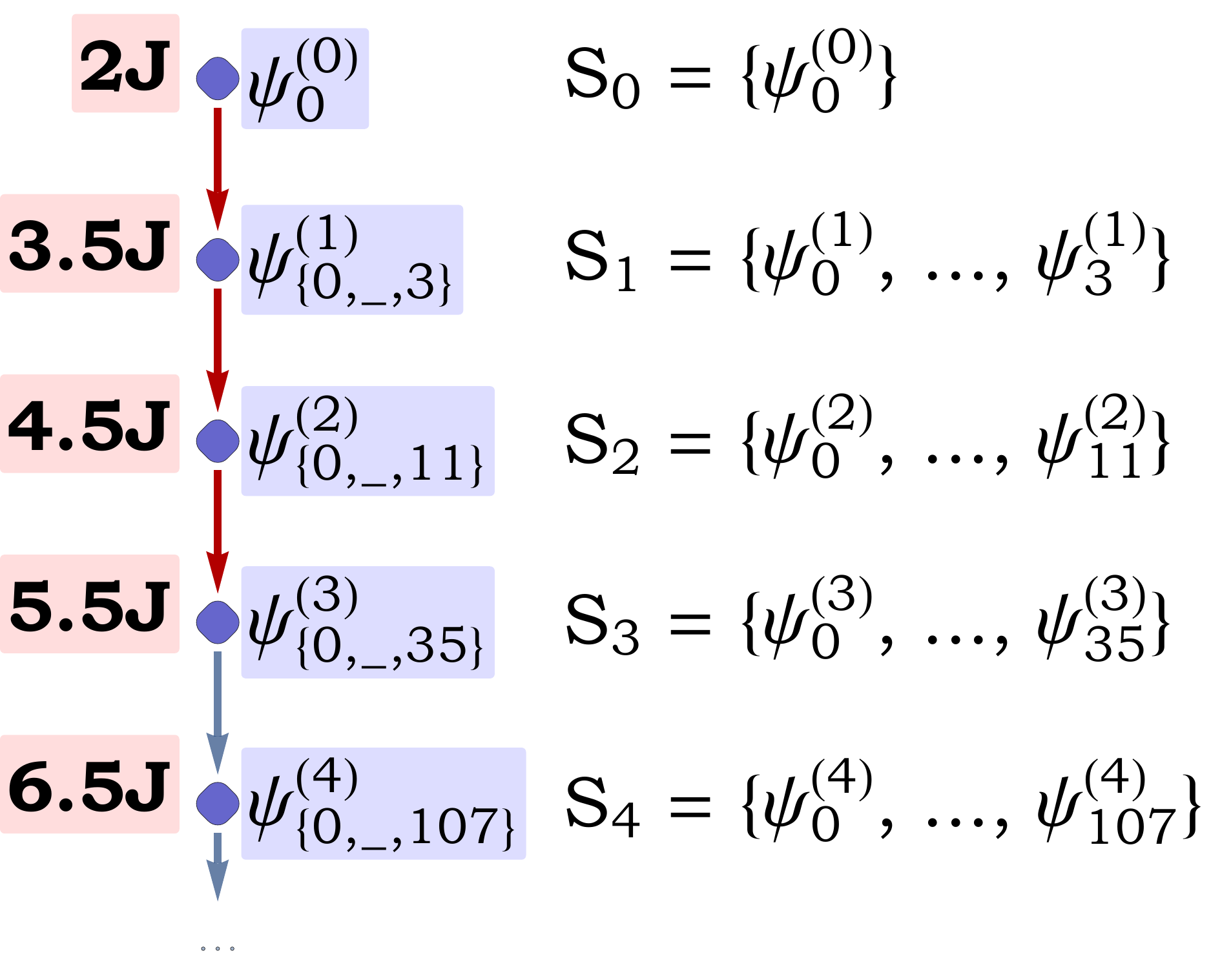}
  \caption{Graph of the reachable states on the Bethe lattice case belonging to the corresponding sets $S_n$. Initial state $\psi_0^{(0)}$ is shown in Fig.~\ref{fig:bethe}.1,. The red arrows correspond to the three hopping processes shown in Figs. \ref{fig:bethe}.1-4. Potential energy of the group of states corresponding to a certain walk (and its symmetries) is denoted next to each node. States with the same number of magnons are indistinguishable. Therefore, a tree-like structure connecting reachable states on the Bethe lattice can be mapped onto a chain.}
    \label{fig:graphbethe}
\end{figure}

Let us start with the Bethe lattice with the coordinate number $z$. Note that then any state $\ket{\psi_{i \in I_n}^{(n)}} \in \mathcal{S}_n$ is an eigenstate of $\mathcal{H}_J$ with the eigenvalue $\lambda_n$. Consider a state $\ket{\psi^{(n)}} = \frac{1}{\sqrt{||\mathcal{S}_n||}}\sum_{i \in I_n} \ket{\psi_i^{(n)}}$ (cf.~Fig.~ \ref{fig:graphbethe}) which is also an eigenstate of $\mathcal{H}_J$,
\begin{equation}
\mathcal{H}_J \ket{\psi^{(n)}} = \lambda_n \ket{\psi^{(n)}}.
\end{equation}
What is more, every term appearing in $-\mathcal{H}_t / t$ is included in the spanning operator $\mathcal{A}^\dag$ or is equivalent to the term in its hermitian conjugate $\mathcal{A}$ when acting on states within $\mathcal{S}_n$. Thus,
\begin{equation} \label{act_Ht}
\mathcal{H}_t \ket{\psi^{(n)}} = b_{n} \ket{\psi^{(n-1)}} + b_{n+1} \ket{\psi^{(n+1)}},
\end{equation}
where $\ket{\psi^{(-1)}} \equiv 0$. In this way we obtain a convenient basis for the states that are reachable by acting with the operator $\hat{G}$,
\begin{equation}
\mathcal{B} = \left\{ \ket{\psi^{(0)}}, \ket{\psi^{(1)}}, \ket{\psi^{(2)}}, ... \right\}.
\end{equation}
In this basis the matrix of the Hamiltonian is tridiagonal,
\begin{equation} \label{ham_matrix}
M(\mathcal{H} - E_\textsc{gs})=\begin{pmatrix}
a_0 & b_1 &   &   &   \\
b_1 &a_1 & b_2 &   &   \\
  & b_2 & a_2 & b_3 &   \\
  &   & b_3 & a_3 & \ddots \\
  &   &   & \ddots & \ddots \\
\end{pmatrix},
\end{equation}
where $ a_n = \lambda_n - E_{\textsc{gs}}$.

In the next step, we calculate all the coefficients of the above-defined Hamiltonian matrix, namely $a_n$ and $b_n$. We start with the latter ones. For $n=1$ we have $||\mathcal{S}_{1}|| = z||\mathcal{S}_{0}||$ thus, following Eq.~(\ref{act_Ht}) and the definition of $\ket{\psi^{(n)}}$, $b_1 = -t\frac{\sqrt{||\mathcal{S}_{1}||}}{\sqrt{||\mathcal{S}_{0}||}} = -t\sqrt{z}$. For $n>1$ it is true that $||\mathcal{S}_{n}|| = (z-1)||\mathcal{S}_{n-1}||$, therefore $b_{n>1} = -t\frac{\sqrt{||\mathcal{S}_{n}||}}{\sqrt{||\mathcal{S}_{n-1}||}} = -t\sqrt{z-1}$.

Calculation of the diagonal coefficients $a_n$ is also rather simple in the case of the Bethe lattice.
A straightforward calculation for $n=0$ yields:
\begin{align}
    a_0 = \frac{zJ}{2}.
\end{align}
Next, we restrict ourselves to $n > 0$ and proceed by mapping $\mathcal{H}_J$ onto a non-interacting operator that has exactly the same spectrum and eigenstates as $\mathcal{H}_J$ assuming there is exactly one hole in the system (which is our case):

First, let us consider a state $\ket{\psi^{(n)}} = \frac{1}{\sqrt{||\mathcal{S}_n||}}\sum_{i \in I_n} \ket{\psi_i^{(n)}}$. There is always a single hole in this state which costs $zJ/2$ due to the term $ \propto h_i^\dag h_i$. Moreover, we can forget about $\propto h_i^\dag h_i h_j^\dag h_j$ terms in $\mathcal{H}_J$ leaving the aforementioned constant $zJ/2$ unaffected. In all considered states with $n>0$ the hole is always in the proximity of exactly one magnon and therefore the energy of each state is lowered by $J/2$ due to the hole-magnon proximity interaction, i.e. the terms $\propto h_i^\dag h_i a_j^\dag a_j$, cf.~Fig.~\ref{fig:bethe}.

Second, for all states with $n > 1$ there is always more than one magnon and the magnon-magnon interactions need to be taken into account. Crucially, for any state $\ket{\psi^{(n)}}$ it is true that magnons form a kind of a one-dimensional chain, i.e. any magnon that is neither the `first' nor the `last' one in such a chain has exactly two magnons as the nearest neighbours, cf.~Fig.~\ref{fig:bethe}. (Note that here we defined: the `first' magnon as the one that is created in place of the hole in state $\ket{\psi^{(0)}}$ once the hole moves to another site and the `last' magnon is the one that is in the proximity of the hole in state $\ket{\psi^{(n)}}$.) The chain-like structure with interactions appearing only on the path of the hole (see~Fig.~\ref{fig:bethe} and~Fig.~\ref{fig:graphbethe}) allows one to reduce the problem to the non-interacting one by substituting,
\begin{equation}
\sum_{\mean{i,j}} 2 a_i^\dag a_i a_j^\dag a_j \rightarrow \frac{1}{z} \sum_{\mean{i,j}}(a_i^\dag a_i + a_j^\dag a_j).
\end{equation}

Altogether, we can write down an equivalent potential energy operator in the following way,
\begin{equation}
\begin{aligned}
\mathcal{H}_J^{\text{equiv}} = E_\textsc{gs} &+ \frac{J}{2} \left(z + 1 - \ket{\psi_0}\bra{\psi_0} \right) \\
&+ J\left(\frac{z}{2}-1\right) \sum_{i} a_i^\dag a_i,
\end{aligned}
\end{equation}
which leads to
\begin{equation}
a_{n>0} = \left( \frac{z + 1}{2} + \left( \frac{z}{2} - 1 \right) n \right) J,
\end{equation}
see also Fig.~\ref{fig:graphbethe}.

Let us stop now for a while and comment on the case of interest in this paper, i.e. $z = 4$, we have 
\begin{align}
a_0 = 2J \quad {\rm and} \quad a_{n>0} = \left( \frac{5}{2} + n \right)J.
\end{align}
This equation can be understood in the following way. Effectively, each magnon costs energy $J$ and a propagating hole costs energy $\frac{5J}{2}$. A static hole, remaining in the position of its creation, effectively gains additional $J/2$ and therefore it costs energy $2J$ with respect to the energy $E_\textsc{gs}$ of the Ising antiferromagnet. In comparison, in the true $\mathcal{H}_J$ both hole and magnon cost $2J$. If there is at least one magnon then the hole-magnon proximity interaction lowers the energy by $J/2$. Every two neighbouring magnons interact gaining energy $J$. We would like to note here that in the case of the square lattice, the argument that magnons form a kind of a one-dimensional chain will no longer be valid and such effective `equivalent potential energy' operator cannot be introduced.

Coming back to the derivation of the local Green's function $G(\omega)$ of a single hole on the Bethe lattice, we note that all we have to do is to calculate one single coefficient of the propagator $\hat{G}$---namely $\bra{\psi_0} \hat{G} \ket{\psi_0}$. Since $\hat{G} = \left( \omega - \mathcal{H} + E_\textsc{gs} \right)^{-1}$ and we know the matrix of the Hamiltonian in basis $\mathcal{B}$, we have
\begin{equation}
G(\omega) = \left[ M(\omega - \mathcal{H} + E_\textsc{gs})^{-1} \right]_{0,0},
\end{equation}
where $[~\cdot~]_{0,0}$ refers to the top left coefficient of the matrix. Now we can partition the matrix,
\begin{equation} \label{eq:matrix_partition}
M(\omega - \mathcal{H} + E_\textsc{gs}) = \begin{pmatrix}
\omega - a_0 & B_1^T  \\
B_1 & \omega - \mathcal{H}_1 + E_\textsc{gs}
\end{pmatrix},
\end{equation}
in order to invert it,
\begin{equation}
\begin{split}
&\left[ M(\omega - \mathcal{H} + E_\textsc{gs})^{-1} \right]_{0,0} = \\ 
&=\left( \omega - a_0 - B_1^T M(\omega - \mathcal{H}_1 + E_\textsc{gs})^{-1} B_1 \right)^{-1} = \\
&=\left( \omega - a_0 - b_1^2 \left[ M(\omega - \mathcal{H}_1 + E_\textsc{gs})^{-1} \right]_{0,0} \right)^{-1}.
\end{split}
\end{equation}
Repeating the procedure for $M(\omega - \mathcal{H}_n + E_\textsc{gs})$ we obtain,
\begin{equation} \label{eq:G_cf}
G(\omega) = \frac{1}{ \displaystyle \omega - a_0 - \frac{b_1^2}{\displaystyle \omega - a_1 - \frac{b_2^2}{\omega - a_2 - \hdots}}}.
\end{equation}
In particular, for $z > 2$, the above continued fraction expansion of $G(\omega)$ can be written in terms of the Bessel functions of the first kind\cite{Che99} leading to the analytic formula,
\begin{equation} \label{eq:G_bethe}
G_{z>2}(\omega) = \left(\omega - \frac{zJ}{2} + \frac{zt}{\sqrt{z-1}}\frac{J_{\Omega(\omega) + 1} \left( 2 \xi \right)}{J_{\Omega(\omega)}\left( 2 \xi \right)} \right)^{-1},
\end{equation}
where
\begin{equation}
\Omega(\omega) = \frac{\frac{z+1}{2} - \frac{\omega}{J}}{ \frac{z}{2}-1}, \quad \text{and} \quad \xi = \frac{t\sqrt{z-1}}{\left(\frac{z}{2} - 1\right)J}.
\end{equation}
For $z = 2$, which is exactly the 1D case, one obtains,
\begin{equation} \label{eq:G_1D}
G_{z=2}(\omega) = \left(\frac{J}{2} \pm \sqrt{\left( \omega - \frac{3J}{2} \right)^2 - 4t^2} \right)^{-1},
\end{equation}
where minus sign $(-)$ in front of the square root applies for $\omega < \frac{3J}{2}$ otherwise plus sign $(+)$ applies. This way we obtain an analytic solution for the local Green's function on the Bethe lattice with the coordinate number $z = 2, 3, 4, ...$.

\subsection{Square lattice}
\begin{figure}[t!]
  \includegraphics[width=0.5\textwidth]{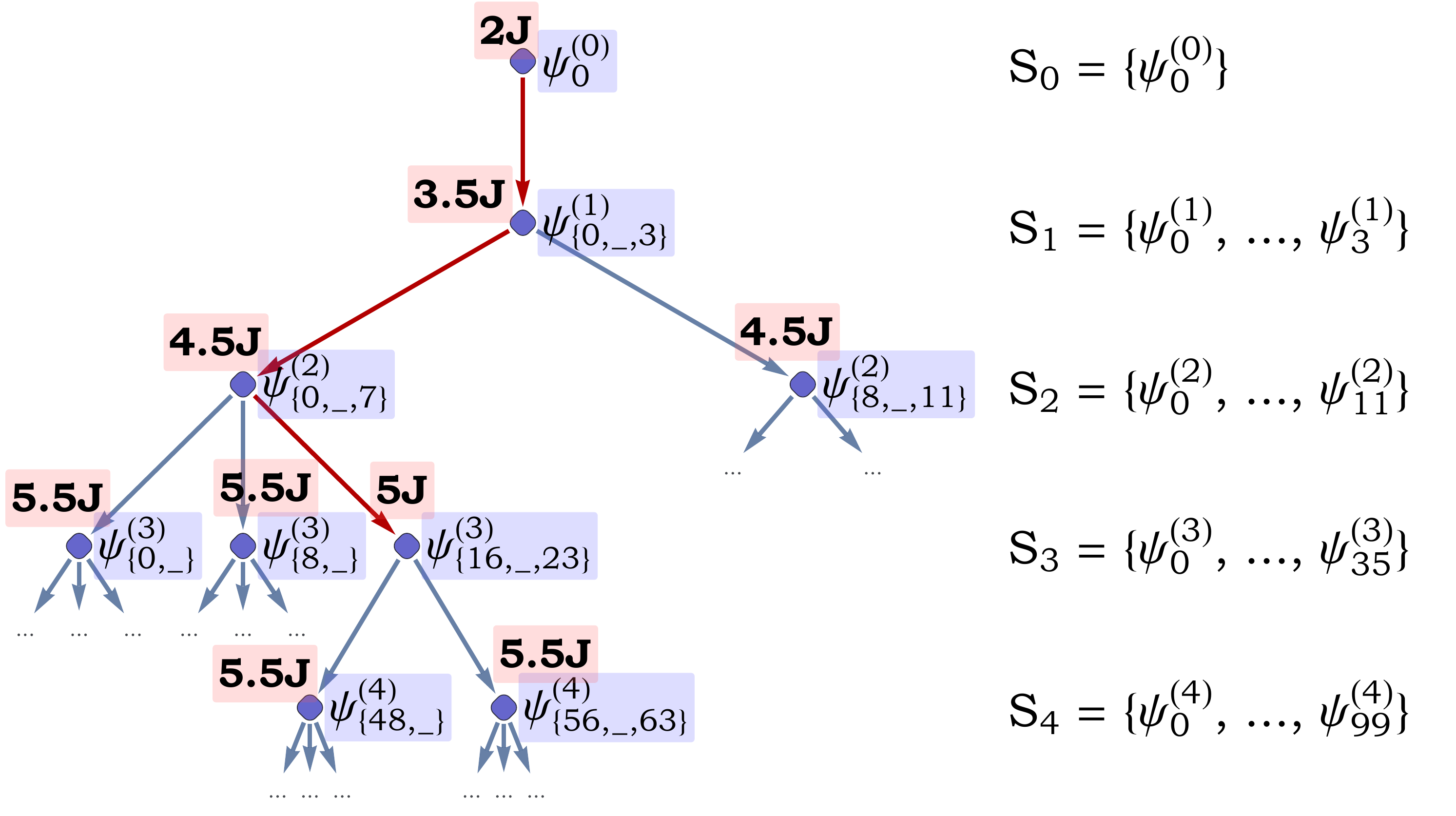}
  \caption{Graph of the reachable states on the square lattice belonging to the corresponding sets $S_n$, within the self-avoiding walks approximation. Initial state $\psi_0^{(0)}$ is shown in Fig. \ref{fig:square}.1,. Red arrows correspond to the three hopping processes shown in Figs. \ref{fig:square}.1-4. Potential energy of the group of states corresponding to a certain walk (and its symmetries) is denoted next to each node. States with the same number of magnons are distinguishable, i.e. they may have different energy. A tree-like structure connecting reachable states cannot be easily mapped onto a chain.}
  \label{fig:graphsquare}
\end{figure}

Even if the matrix of the Hamiltonian is not written in basis $\mathcal{B}$ but in basis $\mathcal{S}$ instead, we still can obtain the desired coefficient of the Greens function by partitioning the matrix in Eq.~(\ref{eq:matrix_partition}). The procedure is then more involved, since the Hamiltonian written in $\mathcal{S}$ is no longer tridiagonal. This is for example the case of the square lattice, see Fig.~\ref{fig:graphsquare}, for the states with the same number of magnons may have different energies due to the magnon-magnon and hole-magnon interaction terms. Thus, {\it a~priori} an analytically closed form for the local Greens function cannot be easily obtained. Nevertheless, we can write down the generic expressions for the Green's function in this case:
\begin{equation} \label{eq:G_SAW}
G(\omega)^{-1} = G_0(\omega)^{-1} - \Sigma_{\psi_0}(\omega),
\end{equation}
where $G_0(\omega)^{-1} = \omega - \omega_{\psi_0}$,
\begin{equation} \label{eq:Sigma_SAW}
\Sigma_{\psi}(\omega) = \sum_{\ket{\phi} \in \mathcal{A}^\dag \ket{\psi}} \frac{t^2}{\omega - \omega_{\phi} - \Sigma_{\phi}(\omega)},
\end{equation}
and $\omega_{\psi} = \bra{\psi} \mathcal{H}_J \ket{\psi}$. The above-defined equation can be understood as a generalization of a continued fraction to a `tree-like' fraction. While the linear form of the continued fraction in Eq.~(\ref{eq:G_cf}) resembles the chain-like graph of Fig.~\ref{fig:graphbethe}, the tree-like structure of Eq.~(\ref{eq:Sigma_SAW}) corresponds to the tree graph of Fig.~\ref{fig:graphsquare}. We use the above-defined equations to calculate the local spectral function $A(\omega)$ for the square lattice within the self-avoiding walks approximation.

\section{\label{sec:results}Results:\\* agreement with ED}

\begin{figure*}[t!]
  \includegraphics[width=1.0\textwidth]{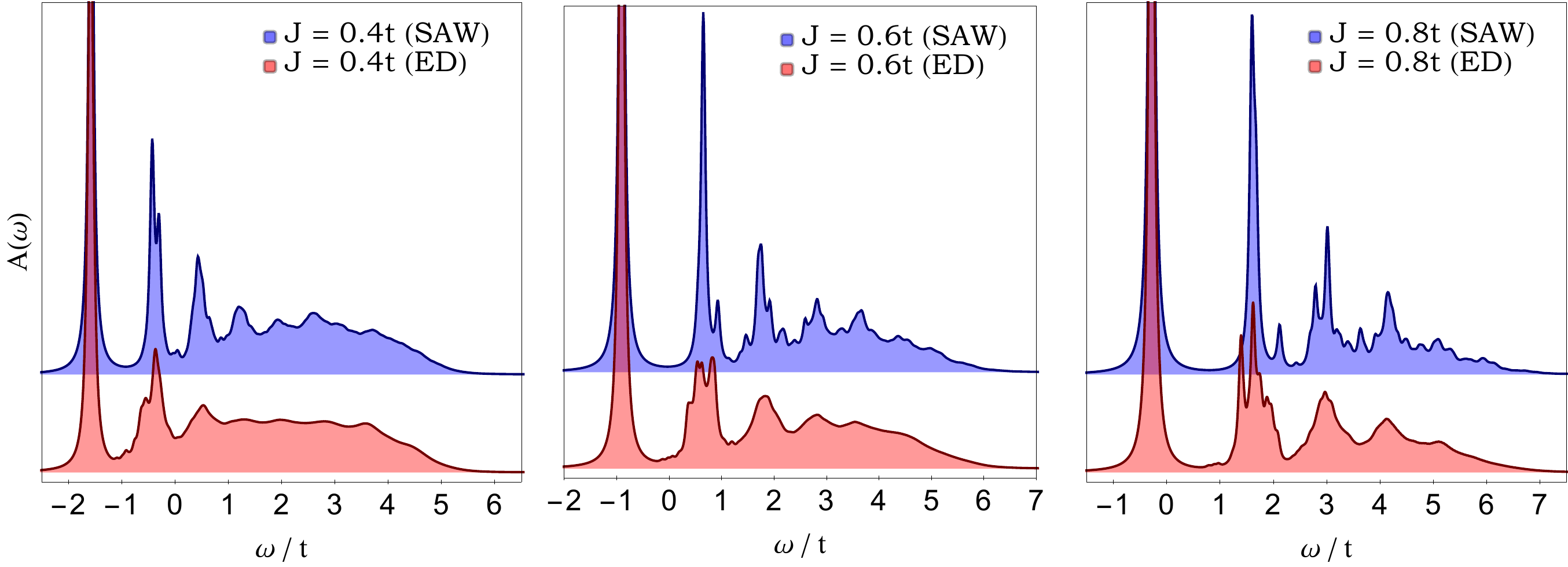}
  \includegraphics[width=1.0\textwidth]{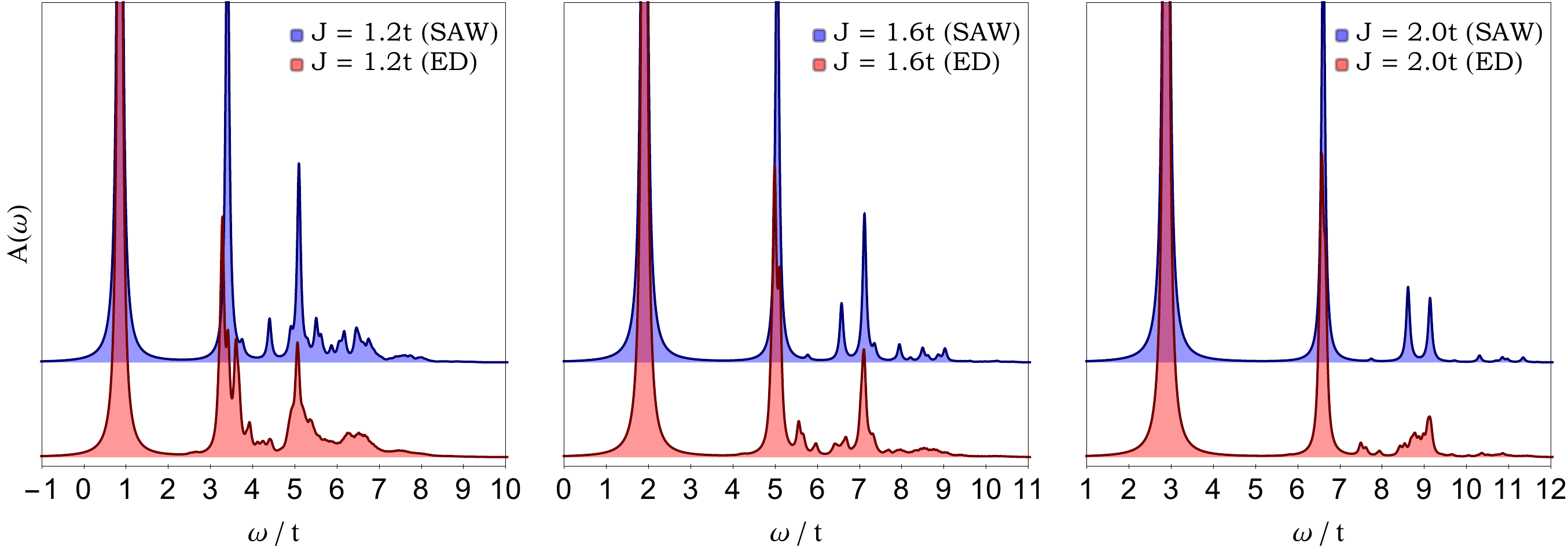}
  \caption{
Spectral function $A(\omega)$ of a single hole in the $t$--$J_z$ model on the 2D square lattice calculated using self-avoiding walks approximation (blue) and ED (red) for $J=0.4t$ (top left), $J=0.6t$ (top middle), $J=0.8t$ (top right), $J=1.2t$ (bottom left), $J=1.6t$ (bottom middle), and $J=2.0t$ (bottom right). Self-avoiding walks approximation includes up to 20 magnons and ED calculations (red) are performed on a 26-site lattice with periodic boundary conditions. Broadening $\delta = 0.05t$.
}
\label{fig:1}
\end{figure*}

In the Methods section (Sec. \ref{sec:methods}) we gave the exact expressions for the single hole Green's function of the \tjz~model on a square lattice in the self-avoiding walks approximation---cf. Eqs. (\ref{eq:G_SAW}-\ref{eq:Sigma_SAW}). As this is not a closed-form expression for the Green's function, but rather a recurrence relation with the nonlinear coefficients, the Green's function has to be found numerically. To this end, we consider all the possible states that have up to 20 magnons, which gives over $1.4 \times 10^9$ of the basis states that are taken into account. This is already enough to obtain converged result of the spectral function $A(\omega)$ for the `canonical' value of the coupling constant $J = 0.4t$. For higher $J$ values even lower number of  magnons is enough. On the other hand, for $J \ll 0.4t$ much larger number of magnons is required to obtain meaningful results. As we are primarily interested in the results around the canonical value of $J/t$ as well as in the benchmarking of the method against the ED, we decided to keep $J \ge 0.4t$ in what follows.

The spectral function $A(\omega)$ calculated for six distinct values of the model parameter $J/t\in [0.4, 2.0]$ is shown in Fig.~\ref{fig:1}. In the limit of realistic of $J<t$ the spectrum consists of a well-separated quasiparticle peak at lowest energy and a relatively large incoherent spectrum. The latter becomes far less pronounced and gains more shape with increasing $J/t$ for $J>t$.

Last but not least, we compare the obtained approximate spectra against the ED results on a finite cluster, cf~Fig.~\ref{fig:1}. (We note that the ED result is obtained on a 26-site cluster and thus suffers from relatively small finite-size effects---as is typical to the \tjz~model w.r.t. the $t$--$J$ model, cf.~\onlinecite{Rie93} and Appendix \hyperref[appendix]{A}). Clearly, the self-avoiding walks approximation gives results qualitatively comparable to the ED calculations for all studied values of $J/t$. In particular, the approximate method reproduces relatively well the overall shape of the spectral function at higher energies. The same applies to the ground state energy and its spectral weight. For the intermediate energies the self-avoiding walks approximation tends to give a more `jagged' spectrum than the ED. 

\begin{figure}[t!]
  \includegraphics[width=0.48\textwidth]{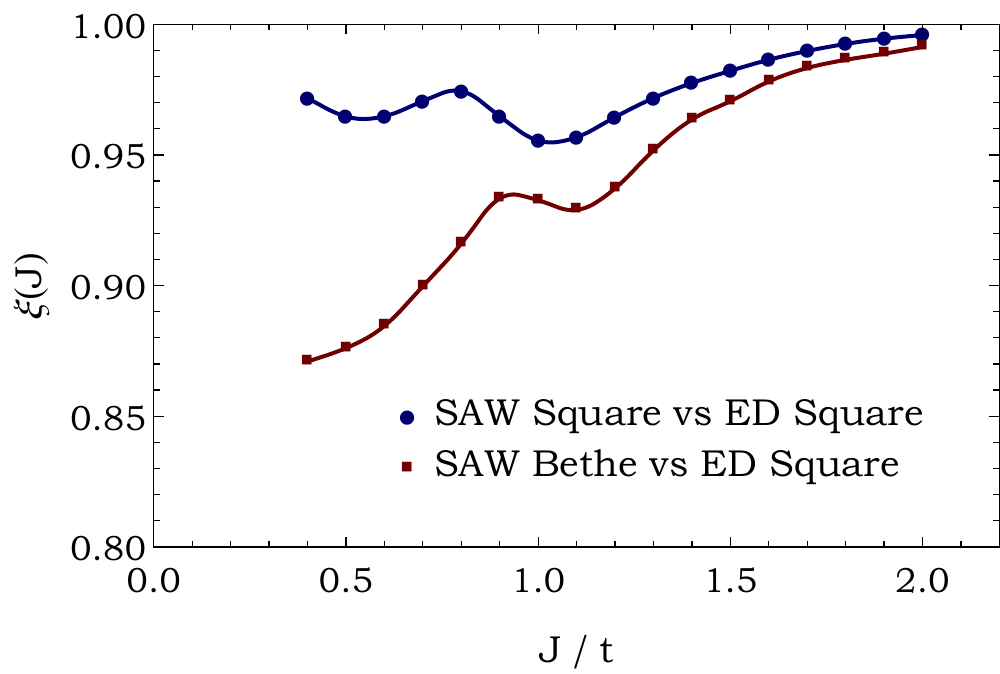}
  \caption{Correlation $\xi(J)$ between the spectral functions $A(\omega)$ calculated on a square and the Bethe lattice within the self-avoiding walks approximation with the ED result on a 26-site square lattice. Note that the self-avoiding walks approximation is an exact method on the Bethe lattice.  
}
\label{fig:overlap}
\end{figure}

To even further substantiate the above claim, we calculate the following correlation function $\xi$ between the two spectra
\begin{equation}
    \xi(A, B) = \frac{\int_{-\infty}^{\infty}A(\omega)B(\omega)\d{\omega}}{\sqrt{\int_{-\infty}^{\infty}|A(\omega)|^2\d{\omega} \int_{-\infty}^{\infty}|B(\omega)|^2\d{\omega}}}.
\end{equation}
The dependence of $\xi$ on the value of the model parameters $J/t$ is shown in Fig.~\ref{fig:overlap}. We observe that indeed the spectrum calculated using the self-avoiding walks approximation on a square and the ED spectrum match very well, as the correlation is always above 95\%---we discuss in Appendix~\hyperref[appendix]{A} the physical origin of this result.  On the other hand, it turns out that a correlation between the self-avoding walks approximation spectrum calculated on a Bethe lattice and the ED result is much worse---especially in the realistic regime of $J<t$. We will come back to the latter result in Sec.~\ref{sec:discussion} in which we explain the origin of the incoherent spectrum in the approximate result on the square lattice.

\begin{figure}[t!]
  \includegraphics[width=0.48\textwidth]{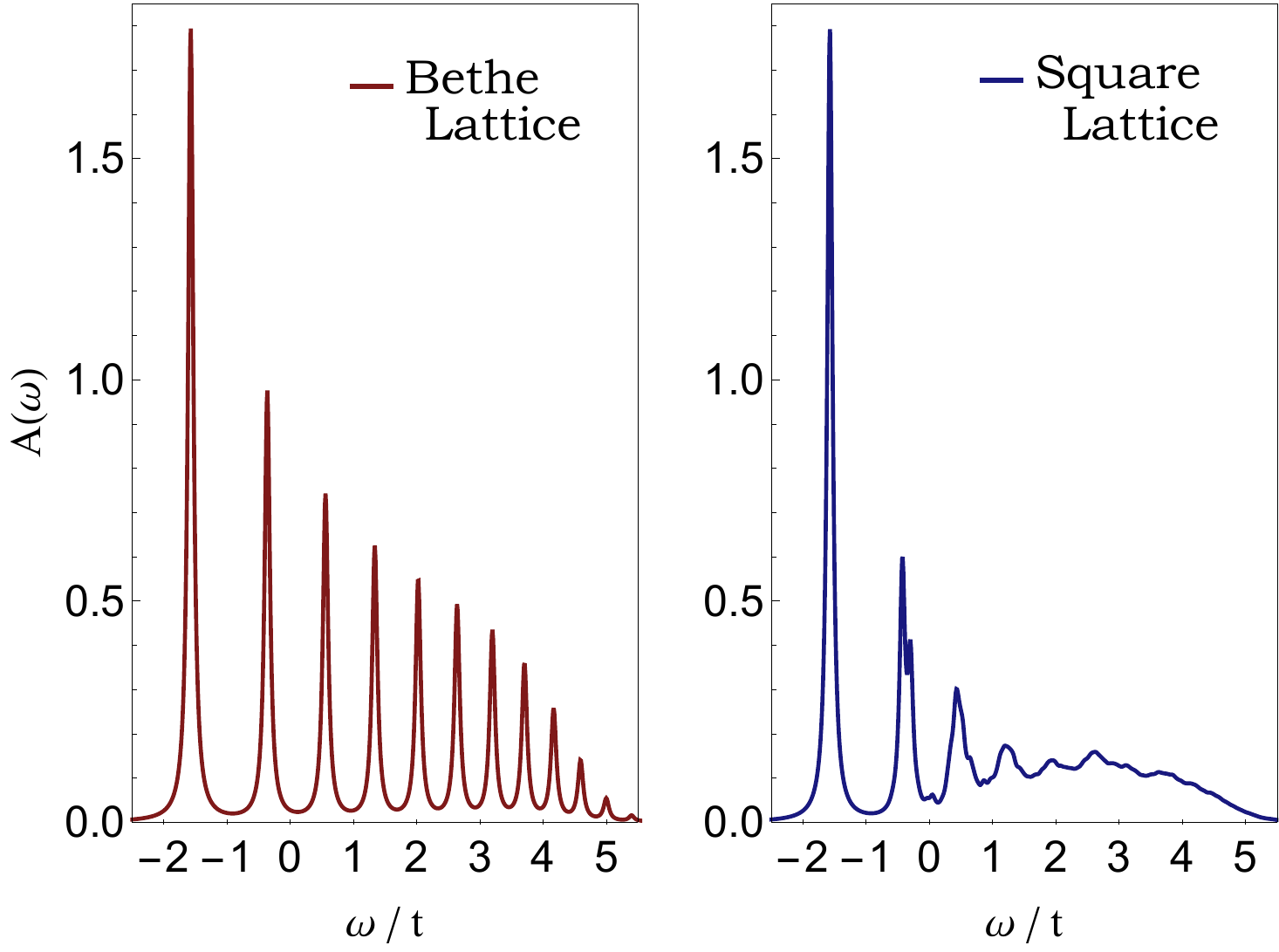}
  \caption{Spectral function $A(\omega)$ of a single hole in the \tjz~model on the Bethe lattice with the coordinate number $z=4$ (left) and the square lattice (right). Results obtained using self-avoiding walks approximation with $J=0.4t$ and broadening $\delta = 0.05t$.
}
\label{fig:2}
\end{figure}

\section{Discussion:\\* origin of the agreement with ED}
\label{sec:agr}

We can address the (surprisingly) small quantitative difference between the approximate spectrum, i.e. calculated on the square lattice using the self-avoiding walks approximation, and the ED spectrum to the loop (Trugman) processes~\cite{Tru88} possible in the ED. This is because the only difference between the self-avoding walks approximation and ED lies in neglecting the paths containing loops. Below we intend to explain in some detail why the loop processes give such a small contribution to the spectrum---in general this is because the number of such paths with loops is relatively small.

Let us first concentrate on the results obtained for large $J/t \in [1, 2]$. In this case the spectrum calculated with ED and with the self-avoiding walks approximation on the square lattice match pretty well (see Fig.~\ref{fig:overlap}, showing the `correlation between the spectral functions' with / without the loops, as well as the spectra for $J/t>1$ of Fig.~\ref{fig:1}). Crucially, the best match is obtained for $J/t=2$ and then this agreement slightly decreases with decreasing $J/t \in [1, 2]$. Such behavior can be best understood by invoking that the lowest order loop corrections to the self-energy scale as $t^6/J^5$, cf. page 321 of~\cite{Mar91}. We stress here that such a simple lowest-order calculation should work in the regime of large $J/t$, since in this case the perturbation theory works~\cite{Mar91} and solely the lowest order corrections should suffice.

On the other hand, the situation encountered for smaller $J/t$ is quite distinct. This is because for intermediate $J/t \in [0.4, 1]$, the agreement between the two methods is still relatively high, i.e. the correlation does {\it not} decrease with decreasing $J/t$ but instead, depending on the value of $J/t$, it oscillates around 95\%-97\% (see Fig.~\ref{fig:overlap}). Here, a note of caution may be in order: the considered correlation function is a very crude measure of the agreement between the two spectra and therefore the observed small changes in the correlation function for intermediate $J/t \in [0.4, 1]$ should rather not be interpreted as pointing towards important changes in the role played by the loops.

We can rationalise the above observation, concerning the  $J/t \in [0.4, 1]$ case, in the following manner. Firstly, one should stress that once $J<t$ one ends up in the strong-coupling limit and hence we should {\it a priori} take into account higher-order corrections in $t/J$. For the paths without loops this means that a `large' number of magnons have to be taken into account in the self-avoiding walks approximation (e.g. for $J/t=0.4$ all states with up to 20 magnons have to be considered compared against 8 magnons for $J/t=2.0$) or that all the `rainbow' diagrams have to be summed over in the SCBA calculations. However, as far as the paths with the loops are concerned, it was suggested in the seminal paper by Trugman~\cite{Tru88}, that it is expected that solely the lowest-order loop correction should be considered---while the higher order loop corrections (i.e. which lead to longer loops) could be neglected, see p. 1599 of~\onlinecite{Tru88}. (While in principle such an important conjecture has to be checked, it requires including the loop processes e.g. in the SCBA calculations and thus is beyond the scope of this paper.) Thus, we encounter here a situation where the paths without loops should be summed to an (almost) infinite order to give reasonable results, though only one type of a path with a loop is {\it probably} relevant. Hence, we expect that, although the contribution of the lowest order loop correction to the self-energy is $\propto t^6/J^5$ and thus grows with decreasing $J/t$, due the relatively small number of loop paths (essentially one type) w.r.t the other paths (essentially infinite), the relevance of the loop paths should {\it not} substantially increase with decreasing $J/t$.

\section{Discussion:\\* Origin of the incoherent spectrum}
\label{sec:discussion}
What is the origin of the substantial incoherent spectrum in the self-avoiding walks approximation result? First, we note that the incoherent spectrum is not triggered by an apparent superposition of the fully coherent (i.e. ladder-like) spectral functions calculated for different momenta and summing up to an incoherent local spectral function. In fact, we have verified that also the momentum-resolved spectral function calculated using ED contains a substantial incoherent spectrum, cf. Appendix \hyperref[appendix]{A}.

Clearly, it is also not triggered by the closed (Trugman) loops~\cite{Tru88}, for the latter are {\it not} included in this approximation as discussed in Appendix~\hyperref[appendix]{A}. Finally, it can also be verified rather easily that the incoherent spectrum is not obtained in the Bethe lattice geometry. While the latter can be easily verified based on the already published results~\cite{Bul68, Kan89, Mar91, Sta96, Che99, Bie19}, to make the paper self-contained we calculate the spectrum for a single hole in the \tjz~model on the Bethe lattice with the coordinate number $z=4$. To this end we make use of the analytic solution for the Bethe lattice case written in Eq.~(\ref{eq:G_bethe}). We remind the self-avoiding walks approximation is exact in this case.

The obtained in this way local spectral function $A(\omega)$ for the Bethe lattice is compared against the corresponding result for the 2D square lattice in Fig.~\ref{fig:2}. Whereas the ground state energy ($-1.57t$ vs $-1.57t$) and its spectral weight (0.2821 vs 0.2815) match extremely well in both cases, the spectrum of the excited states is markedly different. As already discussed in the Introduction (Sec.~\ref{sec:intro}), in the Bethe lattice case the whole spectrum consists of the quasiparticle-like peaks for all energies---a so-called ladder spectrum develops. This is a signature of an (effective) linear potential acting on the mobile hole~\cite{Bul68, Kan89} and suggests that, in the case of the excited states, the single hole on the square lattice case is {\it not exactly} subject to the linear potential.

To investigate the warping of the linear potential on the square lattice, we list the three differences between the hole motion on the Bethe and the square lattice antiferromagnet (all within the self-avoiding walks approximation). {\it First}, the number of sites to which the hole can propagate creating a magnon is different in these two geometries. Whereas on the Bethe lattice (with $z=4$) the number of sites to which the hole can propagate creating a magnon is constant $\mu_{\rm Bethe} = 3$ (apart form the very first step in which case the hole can hop to four sites), for the square lattice this quantity varies with the path of the hole and in the limit of the infinitely long walk tends to the so-called connective constant $\mu_{\rm square} \approx 2.63816$~\cite{Jensen_2003}. However, this difference alone, without considering the interactions discussed below, does not explain the collapse of the ladder spectrum on the square lattice.

The {\it second} and {\it third} difference is far more important. It is related to the hole-magnon proximity interaction (P) as well as the magnon-magnon interaction (M) which differently impacts the hole motion in these two geometries. This is best visible on the cartoon figures showing the hole motion on the square and Bethe lattice geometries, cf.~Figs.~\ref{fig:square}-\ref{fig:bethe}. We observe that P and M interactions may appear along the path of the mobile hole in both lattices but `satellite' (off-path) interactions are possible only in the case of the square lattice. Such `satellite' interactions do not appear in a concerted manner, i.e. not at an equal rate after each hole hopping. In fact, they occur once there exists a site at which the hole path becomes tangential to itself. This means that the linear potential, which is induced by the ever growing number of magnons created along the hole path, becomes warped, for the `satellite' interactions may occur `here and there' (i.e. once the path becomes tangential to itself) when the hole moves on the square lattice.

\begin{figure}[t!]
  \includegraphics[width=0.48\textwidth]{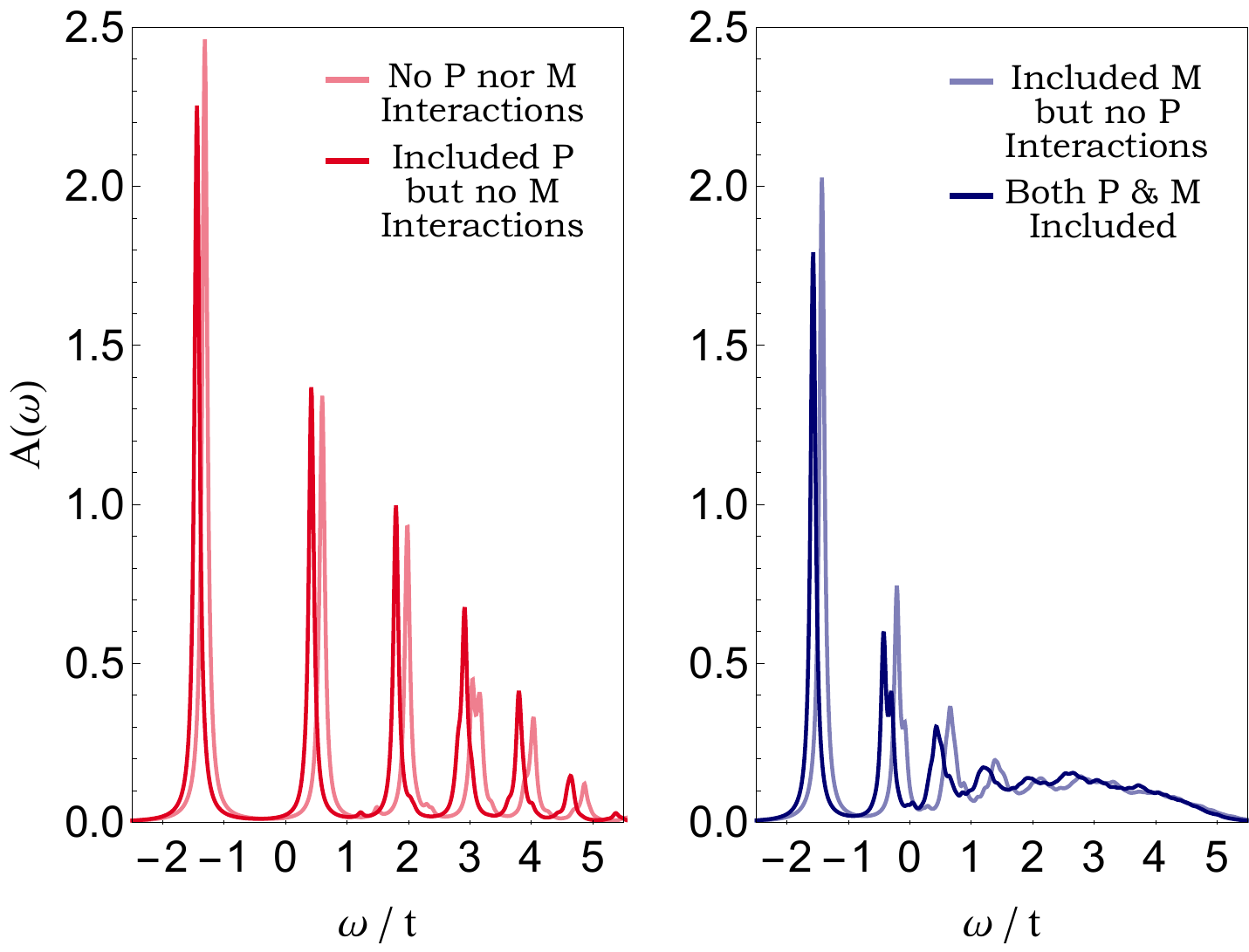}
  \caption{
Spectral function $A(\omega)$ of a single hole in the \tjz model on the square lattice. Left (right) panel shows results without (with) magnon-magnon interactions correctly included in the model Hamiltonian, respectively. Darker (lighter) lines depict results with (without) the hole-magnon proximity interaction properly included, respectively. All results obtained using self-avoiding walks approximation with $J=0.4t$ and broadening $\delta = 0.05t$.
}
\label{fig:3}
\end{figure}

We now test the above conjecture by calculating the spectral function with and without the P as well as the M interactions included in the Hamiltonian, cf.~Fig.~\ref{fig:3}. Clearly, these are the M (magnon-magnon) interactions which are primarily responsible for the onset of the incoherent spectrum. On the other hand, the P (hole-magnon proximity) interactions seem to be far less important in inducing the incoherent spectrum: albeit, on the qualitative level, they also destroy the ladder spectrum, their impact is very small, see Fig.~\ref{fig:3} (left). This can be explained by noting that there is only one single hole in the system while the number of magnons may grow much higher in general and the strength of the interaction between the two magnons is twice as large as the strength of the interaction between the hole and the magnon. Altogether, we conclude that, practically, these are the magnon-magnon interactions  created by the hole moving on the square lattice which are responsible for the onset of the observed incoherent spectrum.

To even further understand the physics related to the motion of the hole along the tangential paths, we discuss how the importance of the tangential paths depends on the model parameters $J/t$. To this end we study the relative contribution of the tangential paths to the overall spectral weight, which can be obtained by subtracting the two correlation functions presented in the Fig.~\ref{fig:overlap}. We observe that, the role of the tangential processes is, in general, gradually suppressed with increasing $J/t$ (except for the small but non-monotonic changes in the correlation function for $J/t \in [0.8, 1.1]$---which, we believe, is largely due to the fact that the correlation function between the two spectra is a very crude measure of the relative role of the contribution of the particular processes to the spectra). Such behavior can be understood in the following way: With increasing $J/t$ the energy cost of a single magnon grows. This holds also for interacting magnons created by the hole moving in a square lattice. At the same time, the longer the hole path is the more energy it costs due to the higher number of magnons for longer paths (even once magnon interactions are included). Thus, with increasing $J/t$ the average length of a path in a particular eigenstate decreases and this pertains to every eigenstate of the problem. Next comes the crucial argument: the shorter the length of such a path is the lower the number of possibilities of the path to be classified as tangential (e.g. paths containing one or two magnons are never tangential). Consequently, the tangential paths should be suppressed with increasing $J/t$--as indeed inferred from Fig.~\ref{fig:overlap}.

\section{\label{sec:conclusions}Conclusions}

\subsection{Summary}
In summary, we compared the spectral function of a single hole in the Ising antiferromagnet on a square and Bethe lattices. Whereas the ground state energy and its spectral weight contribution are almost the same in both cases, the excited parts are qualitatively distinct: unlike for the Bethe lattice the local spectral function on a square lattice is not `ladder-like', i.e. it does not consist of the well-separated quasiparticle-like peaks for any energy. Instead, in the investigated range of $J/t \in [0.4, 2.0]$ the considered spectrum has an important incoherent part (which is substantial in the realistic regime of $J<t$, though it diminishes fast with increasing $J/t$ in the unrealistic limit of $J>t$). The replacement of the nicely-spaced peaks of the ladder-like spectral function by such an incoherent spectrum is due to the warping of an effective linear potential acting on the hole and is primarily attributed to the magnon-magnon interactions. [It is {\it not} related to the hole moving along the (Trugman) loops~\cite{Tru88} for the latter processes lead to a relatively smaller number of the allowed paths, give a much smaller contribution to the spectral function, and therefore, albeit non-negligible, can firstly be neglected, cf.~Sec.~\ref{sec:agr}.]
 
The easiest way to understand this result is to consider that once the hole moves on a lattice the cost of the magnons created along the path of the hole can partially be lowered due to the magnon-magnon interactions (i.e. the local antiferromagnetic correlations can be reconstructed). Crucially, on the square lattice magnons interact not only along the hole path but also on the bonds connecting the tangential points of the path via the so-called `satellite' interactions (cf.~Fig~\ref{fig:square}). 

By definition such `satellite' interactions cannot take place on a Bethe lattice, since in this case the hole creates magnons along an `isolated' chain and therefore the magnon-magnon interaction just leads to a shift in the energy of the created magnons. Consequently, unlike in the case of a square lattice, the magnon-magnon interactions on the Bethe lattice do not remove the degeneracy of the eigenstates of the $t$--$J_z$ model with a single hole and the exact spectral function is always ladder-like~\cite{Bul68, Kan89, Mar91, Shr88, Sta96, Che99, Bie19}.

\subsection{Outlook}\label{sec:outlook}
In general, this study shows that the, often forgotten, magnon-magnon interactions affect the physics of the doped $t$--$J^z$ model quite drastically even in 2D (that such interactions are important in 1D is shown e.g. in Ref.~\onlinecite{Bie19}). Naturally, it is interesting to ask what could be the impact of the magnon-magnon interactions for the motion of a single hole in the half-filled $t$--$J$ model in 2D, as relevant for the studies of e.g. the doped copper oxides (although, interestingly, the extended \tjz~model might be enough for the cuprates~\cite{Ebr14}). While a detailed understanding of this problem is beyond the scope of this work, we suggest that:

(i) For the {\it Bethe lattice} the role of the magnon-magnon interactions in the motion of a hole in the $t$--$J$ model is probably only quantitative, since (as suggested by this paper) the magnon-magnon interactions would merely `rescale' the string potential---and the latter is anyway partially `erased' by the spin flip processes of the $t$--$J$ Hamiltonian, cf.~Ref.~\onlinecite{Manousakis2007}.

(ii) For the {\it square lattice} adding the magnon-magnon interactions to the $t$--$J$ model problem treated on the linear spin wave approximation level might in principle lead to some qualitative differences. In fact, in this case it would be both the magnon-magnon interactions as well as the spin flip terms which should `help' in reducing the strings created by the hole. Thus, we expect that the hole should be able to move even `more easily' and that the spectrum should be even less ladder-like in this exact case than in the case of the 2D t-J model calculated using the linear spin wave approximation.

We also would like to point out a possible application of the self-avoiding walks approximation to those numerical methods, which are currently subject to the Bethe lattice approximation (but should rather be calculated on hypercubic lattices). Moreover, one could think of a generalization of this method to the more abstract spaces, such as the Hilbert space itself, in order to \textit{track} the desired quantities of a given system, similarly to the addressed in this paper role of the magnon-magnon interactions in the \tjz~model with a single hole.

Finally, as a side message, the paper also shows: (i) that the differences between the Bethe and square lattices should not be disregarded; and (ii) how rather subtle, and often neglected terms in the Hamiltonian, can completely alter the neat quasiparticle-like behavior and lead to the `unparticle-like' (incoherent) physics.

\textit{Note added in proof.} Recently we got aware of two recent studies which support some of the
findings and suggestions of this paper :
(i) Results of paper~[\onlinecite{PhysRevX.8.011046}], {\it inter alia}, show that indeed the role
of the Trugman loops in the $t$--$J^z$ model
is relatively small---in agreement with the results of Sec.~\ref{sec:agr} of this paper;
(ii) Results of paper~[\onlinecite{PhysRevB.102.035139}], {\it inter alia}, show that indeed the
higher vibrational peaks are remarkably absent
in the $t$--$J$ model spectra---in agreement with the suggestions of
Sec.~\ref{sec:outlook} where we write that
any signatures of the ladder spectrum should be very weak in the $t$--$J$ model.

\section{Acknowledgments}
We thank Krzysztof Bieniasz, Andrzej M. Ole\'s and Yao Wang for stimulating discussions. We kindly acknowledge support by the (Polish) National Science Center (NCN, Poland) under Projects No. 2016/22/E/ST3/00560 (PW and KW) and 2016/23/B/ST3/00839 (KW).

\appendix \label{appendix}

\section{Momentum resolved spectral function}
In order to show that the obtained in ED {\it incoherent local} spectral function is {\it not} triggered by an apparent superposition of the {\it coherent momentum-dependent} spectral functions calculated for different momenta (which would sum up to an incoherent local spectral function), we calculate the momentum resolved spectral function $A_\sigma(k,\omega)$ of the single hole doped to the Ising antiferromagnet. The latter is defined as
\begin{equation}
    \begin{aligned}
    &A_\sigma(k,\omega) = -\frac{1}{\pi}\lim_{\delta\to0^+} \text{Im}\{G_\sigma(k, \omega + i\delta)\}, \\
    &G_\sigma(k, \omega) = \Bra{\textsc{gs}} \tilde{c}_{k,\sigma}^\dag \frac{1}{\omega - \mathcal{H} + E_\textsc{gs}} \tilde{c}_{k,\sigma} \Ket{\textsc{gs}},
    \end{aligned}
\end{equation}
where $E_\textsc{gs}$ stands for the ground state energy (and otherwise the notation as in the main text of the paper).
\begin{figure*}[t!]
  \includegraphics[width=1.0\textwidth]{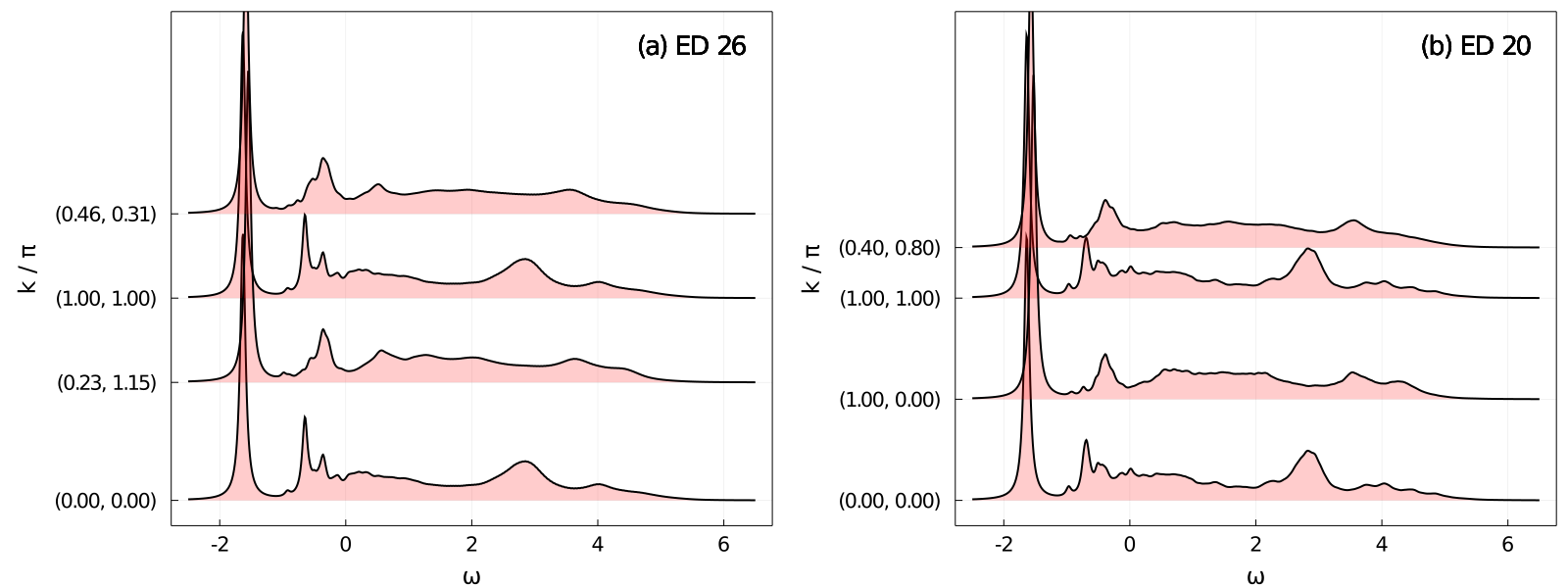}
   \caption{Spectral function $A(k,\omega)$ of a single hole in the \tjz~model on the square lattice. Results obtained using exact diagonalisation (ED) on a 26-site [panel $(a)$] and 20-site [panel $(b)$] square lattice with $J=0.4t$ and broadening $\delta = 0.05t$.}
  \label{fig:6}
\end{figure*}
The results were obtained for the 20- and 26-sites square lattice using the ED (Lanczos) method. Crucially, for all momenta $k$ a continuum of states can be observed in the calculated spectral function, see Fig.~\ref{fig:6}. This suggests that indeed the incoherent spectrum obtained in ED is {\it not} formed by a superposition of the completely coherent (i.e. ladder-like) $A_\sigma(k,\omega)$. Moreover, these results also show the relatively small finite size effects of the 26-site cluster (as the differences between the spectrum obtained on 20 and 26-sites are merely quantitative and rather small). Finally, as a side note, we also confirm a rather low momentum dependence of the ground state originating in the Trugman processes~\cite{Tru88}, i.e. when the hole is allowed to walk along the loops.

\bibliographystyle{apsrev4-1}

\end{document}